\documentclass[prb,twocolumn,twocolumngrid]{revtex4}
\usepackage{graphicx}
\usepackage{amsfonts}
\usepackage{amsmath}
\usepackage{bm}
\usepackage{alltt}
\usepackage{fancyhdr}
\begin{document}

\title{Geometry Optimization of Crystals by the Quasi-\\
       Independent Curvilinear Coordinate Approximation }

\author{K\'aroly N\'emeth}
\email{Nemeth@ANL.Gov}
\author{Matt Challacombe}
\affiliation{Theoretical Division,\\ Los Alamos National Laboratory, \\ Los Alamos, NM 87545, USA}

\date{\today}

\begin{abstract}
The quasi-independent curvilinear coordinate approximation (QUICCA) method 
[K. N\'emeth and M. Challacombe, J. Chem. Phys. {\bf 121}, 2877, (2004)] 
is extended to the optimization of crystal structures.  We demonstrate that QUICCA 
is valid under periodic boundary conditions, enabling simultaneous relaxation of the 
lattice and atomic coordinates, as illustrated by tight optimization of 
polyethylene, hexagonal boron-nitride, a (10,0) carbon-nanotube, hexagonal ice,
quartz and sulfur at the $\Gamma$-point RPBE/STO-3G level of theory.
\end{abstract}


\maketitle

\preprint{LA-UR 05-4015}

\section{Introduction}
Internal coordinates, involving bond stretches, angle bends, torsions and out of plane bends, etc. 
are now routinely used in the optimization of 
molecular structures by most standard quantum chemistry programs. 
Internal coordinates are advantageous for geometry optimization as 
they exhibit a reduced harmonic and anharmonic vibrational coupling 
\cite{PPulay69,GFogarasi79,GFogarasi92,PPulay77}.
This effect allows for larger steps during optimization, reducing the 
number of steps by 7-10 times for small to medium sized molecules, relative 
to Cartesian conjugate gradient schemes \cite{TBucko05}.

Recently, Kudin, Scuseria and Schlegel (KSS) \cite{KKudin01} and Andzelm, King-Smith and Fitzgerald (AKF) 
\cite{JAndzelm01}
reported the first crystal structure optimizations using internal coordinates.
These authors proposed new ways of building Wilson's B matrix \cite{EWilson55} 
under periodic boundary conditions.  The B-matrix (and its higher order generalizations), 
defines the transformation between energy derivatives in Cartesian coordinates and those 
in internal coordinates \cite{EWilson55,MChallacombe91}. 
The scheme presented by KSS allows the simultaneous relaxation of atomic positions and lattice parameters,
while the method developed by AKF allows relaxation of atomic positions with a fixed lattice.

More recently still, Bu\v{c}ko, Hafner and {\'A}ngy{\'a}n (BHA) \cite{TBucko05}
presented a more ``democratic'' formulation of Wilson's B matrix for crystals, 
realizing that changes in periodic internal coordinates may have non-vanishing 
contributions to the lattice.  The BHA definition of the B-matrix will be used 
throughout this paper. 

In the present article, we extend our recently developed optimization algorithm, 
the Quasi Independent Curvilinear Coordinate Approach (QUICCA) \cite{KNemeth04} 
to the condensed phase.  QUICCA is based on the idea  that optimization 
can be carried out in an uncoupled internal coordinate representation, with coupling 
effects accounted for implicitly through a weighted fit. So far,
QUICCA has been simple to implement, robust and efficient for isolated 
molecules, with a computational cost that scales linearly with system size.
However, nothing is yet known about the applicability of QUICCA for crystals. 
In crystals, coupling  may be very different from that encountered in isolated 
molecules, and there may be significant effects from changes in the lattice.
If QUICCA also works well for crystals, then it presents a viable offshoot 
of gradient only algorithms for the large scale optimization of atomistic 
systems. 

In the following, we first review the construction of Wilson's B matrix
for crystals (Sec.~\ref{crystalBmat}), and then in Sec.~\ref{implementation}
we go over the QUICCA algorithm  and discuss its implementation for 
periodic systems.  In Sec.~\ref{results}, results of test calculations 
ranging over a broad class of crystalline systems are presented.
We discuss these results in Sec.~\ref{discussion}, and then go on to 
present our conclusions in Sec.~\ref{conclusion}.

\section{Methodology}

\subsection{Wilson's B matrix for crystals} \label{crystalBmat}

In this section we briefly review the BHA approach to construction
of the periodic B-matrix, with a somewhat simplified notation.  
A more detailed discussion can be found in Ref.~[\onlinecite{TBucko05}] 
and a background gained from Wilson's book, Ref.~[\onlinecite{EWilson55}].

The independent geometrical variables of a crystal are the 
fractional coordinates $\vec{f}_{k}$, and the lattice vectors $\vec{h}_{l}$.
The absolute Cartesian position of the $k$-th atom, $\vec{r}_{k}$, 
is related to the corresponding fractional coordinates by
\begin{equation}
\vec{r}_{k} = {\bf h} \vec{f}_{k} ,
\end{equation}
where ${\bf h}=\{\vec{h}_1:\vec{h}_2:\vec{h}_3\}$ is the matrix of (coloumn wise) Cartesian lattice vectors.

The periodic B-matrix is naturally coloumn blocked, with a fractional coordinate block 
${\bf B}^f_{\scriptscriptstyle N_i \times 3 N_a }$,
and a lattice coordinate block ${\bf B}^h_{\scriptscriptstyle N_i \times 9}$.  The full B-matrix
is then ${\bf B} = \left( {\bf B}^f : {\bf B}^h \right)$, a $N_i \times (3 N_a +9)$ matrix
with $N_i$ the number of internal coordinates and $N_a$ the number of atoms.  
Elements of ${\bf B}^f$ involve total derivatives of internal coordinates with respect to 
fractionals,
\begin{equation}
\vec{b}_{ik} = \frac{d \phi_{i}}{d \vec{f}_{k}} , \qquad k=1,N_a \, ,
\end{equation}
where $\phi_{i}$ is the $i$-th internal coordinate and the B-matrix elements are atom-blocked as the 
three-vector $\vec{b}_{ik}$.  Only the atoms $k$ that determine internal coordinate $i$ are 
non-zero, making ${\bf B}^f$ extremely sparse.  Conventional elements of Wilson's B-matrix, 
$\frac{\partial \phi_{i}}{\partial \vec{r}_{k}}$, are related to these total derivatives by the 
chain rule:
\begin{equation}
\vec{b}_{ik} = \frac{d \phi_{i}}{d \vec{f}_{k}} = \frac{\partial \phi_{i}}{\partial \vec{r}_{k}} {\bold h}.
\label{Bfractional}
\end{equation}
Likewise, elements of ${\bf B}^h$ are 
\begin{equation} \label{Blattice}
\vec{b}_{il} = \frac{d \phi_{i}}{d h_{l}} = \sum_{m \in {\bf n}_i} 
          \frac{\partial \phi_{i}}{\partial \vec{r}_{m}} f_{ml}, \qquad l=1,3
\end{equation}
where $f_{ml}$ is the $l$-th component of the fractional coordinate corresponding to 
atom $m$, and the summation goes over the set ${\bf n}_i $ of all atoms that determine $\phi_{i}$,
i.e. in case of torsions and out of plane bendings, there are four atoms in this set. 
From Eq.~(\ref{Blattice}), it is clear that change in the lattice has the potential to change 
each internal coordinate, through the fractionals and depending on symmetry, so that in 
general we can expect ${\bf B}^h$ to be dense.  Also, note that $f_{ml}$ can be greater then 
$1$ if atom $m$ is not in the central cell. 

\subsection{Internal Coordinate Transformations}

With the B-matrix properly defined, transformation of Cartesian coordinates, lattice vectors  and their 
corresponding gradients into internal coordinates remains.  As in the case of the B-matrix, the 
Cartesian gradients are partitioned into a fractional component,
\begin{equation}
\vec{g}^{\,f}_k = \frac{d E}{d \vec{f}_{k}} 
= \frac{\partial E}{\partial \vec{r}_{k}} {\bf h}, \qquad k=1,N_a \, ,
\end{equation}
and a lattice component, $\vec{g}^h$, with 9 entries involving the {\em total derivatives}
$\frac{d E}{d \vec{h}_{l}}$; the emphasis underscores the fact that some
programs produce partial derivatives with respect to lattice vectors, and this subtly
must be taken into account (e.g. see Eq.~(9) in Ref.~[\onlinecite{TBucko05}]).  

With this blocking structure, gradients in internal coordinates are then defined implicitly by the equation
\begin{equation}
\vec{g}^{\, i} \left( {\bf B}^f : {\bf B}^h \right)= \left(\vec{g}^{\, f} : \vec{g}^{\, h} \right) \;, 
\end{equation}
which  may be solved in linear scaling CPU time, through Cholesky factorization of the 
matrix ${\bf B}^{t} {\bf B}$ that enters the left-handed pseudo-inverse, 
followed by forward and backward substitution, as described in Ref.~[\onlinecite{KNemeth00B}].  
Linear scaling factorization of ${\bf B}^t \bf B$ may be achieved, as the upper left-hand 
$3 N_a \times 3 N_a$ block is hyper-sparse, leading also to a hyper-sparse elimination tree
and sparse Cholesky factors, similar to the case of isolated molecules \cite{KNemeth00B}.
Note however that the factorization of ${\bf B B}^{t}$, involved in construction of a right-handed 
pseudo-inverse, has the dense $9 \times N_i$ row uppermost and the dense $N_i \times 9$ coloumn foremost,
leading to dense Cholesky factors. Thus, the left- and right-handed approach  are not equivalent for 
internal coordinate transformations of large crystal systems, as they are in the case of
gas phase molecules, when using sparse linear algebra to achieve a reduced computational cost.

Once predictions are made for improved values of internal coordinates,
based on the internal coordinates and their gradients, new Cartesian coordinates are 
calculated via an iterative back-transformation as described in Ref.~[\onlinecite{KNemeth04}], 
which also scales linearly with the system-size. At each step of the iterative back-transformation, 
fractional coordinates 
and lattice vectors are updated and the corresponding atomic Cartesian coordinates are
computed.  From these, updated values of the internal coordinates are found, and
the next iteration is started.  Besides these minor details, the back-transformation
is the same as that used for isolated molecules.

\section{Implementation} \label{implementation}

\begin{figure}[h]
\resizebox*{3.5in}{!}{\includegraphics{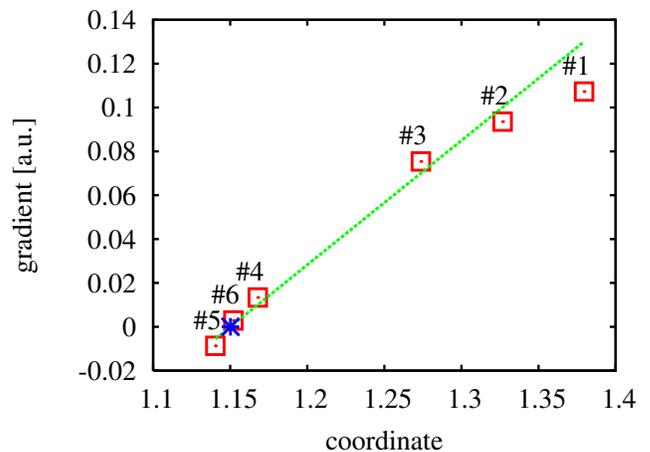}}
\caption{Progression of RHF-MIC/STO-3G gradients for a hydrogen bond coordinate
in hexagonal ice, starting from the crystal structure \cite{AGoto90}.  
Numbers label QUICCA optimization steps, and the star shows the predicted 
minimum  given by a weighted line fit (dashed line) of the previous 6 points.}
\label{iceIh}
\end{figure}

\subsection{The QUICCA algorithm for crystals} 

Details of the QUasi-Independent Curvilinear Coordinate Approximation 
(QUICCA) for geometry optimization of isolated molecules have been 
given in Ref.~[\onlinecite{KNemeth04}].  Here we provide a brief overview 
of the method.  

QUICCA is based on the trends shown by internal coordinate 
gradients during geometry optimization.  See Fig.~\ref{iceIh} for example.
These trends can be exploited by a weighted curve fit for each internal 
coordinate gradient,  allowing an independent, one-dimensional extrapolation 
to zero. The predicted minima for each internal coordinate  collectively 
determines one optimization step.  This method works surprisingly well, but 
only when the weights are chosen to account for coupling effects;  when 
coupling is strong, the weights should be small and vica versa.  Also, the
fitting process has an important averaging effect on coupling that contributes 
to the success of QUICCA.

The only difference in our current implementation of QUICCA, relative
to that described in Ref.~[\onlinecite{KNemeth04}], is that merging 
the connectivities from recent optimization steps is no longer carried out.
Omitting this connectivity merger does not change the results for Baker's 
test suite as given in Ref.~[\onlinecite{KNemeth04}], and does not appear
to diminish the overall effectiveness of QUICCA.

\subsection{The periodic internal coordinate system}

In setting up a periodic internal coordinate system, it is first 
important to consider the situation where, during optimization, atoms 
wrapping across cell boundaries lead to large (unphysical) jumps in 
bond lengths, angles, etc.  This situation is avoided here by employing a minimum 
image criterion to generate a set of Cartesian coordinates consistent 
with a fixed reference geometry.  

Also, because internal coordinates span cell boundaries, it is 
convenient to work with a $3\times3\times3$ supercell,
including the central cell surrounded by its 27 nearest neighbors.
Even though a smaller replica of 8 cells, with lattice indices
between $0$ and $1$ contains all necessary local internal coordinates,
we prefer to employ the larger supercell, to avoid fragmentation
of bonds etc at the cell boundaries.  Then, all internal coordinates 
are identified in the supercell by means of a recognition algorithm,
just as for isolated molecules.  Finally, internal coordinates are discarded 
that do not involve at least one atom in the central cell. 

This procedure produces symmetry equivalent internal coordinates, 
among those internal coordinates that cross cell boundaries.
In the present implementation these equivalent coordinates are not 
filtered out, since their presence has no major effect on the optimization;
the equivalent coordinates result in exactly the same line fit and same
predicted minima.

It is worth noting that an appropriate internal coordinate 
recognition scheme is extremely important to the success of internal 
coordinate optimization.  Here, we are using a still experimental algorithm, 
which we hope to describe in a forthcoming paper.

\subsection{Treatment of constraints}

In the treatment of  constraints, we distinguish between soft and hard constraints.
Soft constraints approach their target value as the optimization proceeds,  reaching it at 
convergence.  Most internal coordinate constraints are of the soft type in our implementation. 
The application of soft constraints is particularly useful in situations where it is difficult
to construct corresponding Cartesian coordinates that satisfy the constrained values.
Hard constraints are set to their required value  at the beginning and keep their value during 
the optimization.  Cartesian  and lattice constraints are hard constraints in the current 
implementation. 

Our method of treating hard constraints is similar to Baker's projection scheme \cite{JBaker96};  
columns of the B-matrix corresponding to hard constraints are simply zeroed. 
This zeroing reflects the simple fact that a constrained Cartesian coordinate or
lattice parameter may not vary any internal coordinate.
Note, that if the lattice parameters $a, b, c, \alpha, \beta$ and $\gamma$ are constrained, 
${\bf B}^h$ must be transformed from a lattice-vector representation into a lattice-parameter 
representation by using the generalized inverse of the lattice-parameter Jacobian. 
This transformation results in 6 columns corresponding to the 6 lattice parameters. 
After zeroing the relevant columns, this portion of the ${\bf B}^h$ matrix is transformed back 
into the original $N_i \times 9$ representation.   In addition, this approach
guarantees that during the iterative back-transformation no displacements occur 
for constrained coordinates or parameters.

In both cases, it is necessary to project out the constraint-space component of the 
Cartesian gradients, so that the internal coordinate gradients remain consistent with 
the constrained internal coordinates. As recommended by Pulay \cite{PPulay77}, this is accomplished 
via projection;
\begin{equation}
\vec{g}\;' = {\bf P} \vec{g} \, , \label{initialp}
\end{equation}
where $\bf P$ is the purification projector that filters out the constraint-space.
For Cartesian variables, this together the aforementioned zeroing of ${\bf B}^f$ is sufficient 
to enforce a hard constraint.  For constrained internal coordinate variables though, this
projection is not entirely sufficient, as there are further contaminants that can arise in
the left-handed pseudo-inverse transformation to internal coordinates.  These contaminants
can be rigorously removed by introducing a further projection in the transformation step, 
as suggested by Baker \cite{JBaker96}.
However, toward convergence these contaminants disappear, and in practice, we find good 
performance without introducing an additional purification step.  And so, Eq.~(\ref{initialp})
is the only purification used in the current implementation, for both hard and soft constraints.

While purification of the gradients is sufficient for the already satisfied hard constraints,
the soft internal coordinate constraints must be imposed at each geometry step.  This is 
accomplished through setting the constrained and optimized internal coordinates, followed
by iterative back transformation to Cartesian coordinates.  This procedure finds a closest 
fit that satisfies the modified internal coordinate system, as described in 
Refs.~[\onlinecite{PPulay77}] and [\onlinecite{KNemeth00B,KNemeth01}].

\subsection{Implementation}

Crystal QUICCA has been implemented in the MondoSCF suite of linear scaling 
quantum chemistry codes \cite{MondoSCF}, using FORTRAN-95 and sparse (non-atom-blocked) 
linear algebra.  Total energies are computed using existing fast methods 
(TRS4\cite{ANiklasson03}, ONX \cite{CTymczak05a}, QCTC and HiCu \cite{CTymczak05b}), 
and the corresponding lattice forces 
(total derivatives) are calculated analytically, with related methods that will be 
described eventually. Full linear scaling algorithms have been used throughout.

These linear scaling algorithms deliver $\Gamma$-point energies and forces only.  
For the Hartree-Fock (HF) model, this corresponds to the minimum image criterion 
(HF-MIC) \cite{CTymczak05a}.  For small unit cells, these $\Gamma$-point effects typically lead 
to different values for symmetry equivalent bond and lattice parameters.  These effects 
decay rapidly with system size, and are typically less severe for pure DFT than for HF-MIC.  

While not always the most efficient option, backtracking has been used in all calculations.
Backtracking proceeds by reducing the steplength by halves, for up to three cycles.  
After that, QUICCA accepts the higher energy and carries on.  

In all calculations, the {\tt TIGHT} numerical thresholding scheme \cite{CTymczak05a} has been used, 
targeting a relative error of {\tt 1D-8} in the total energy and an 
absolute error of {\tt 1D-4} in the forces.  A single convergence criterion is used.
That criterion is that the maximum magnitude of both atomic and lattice vector gradients 
is less than {\tt 5D-4} au at convergence.

Atomic units are used throughout.

\section{Results} \label{results}

\subsection{Test set}


\begin{figure}[h]
\resizebox*{3.5in}{!}{\includegraphics{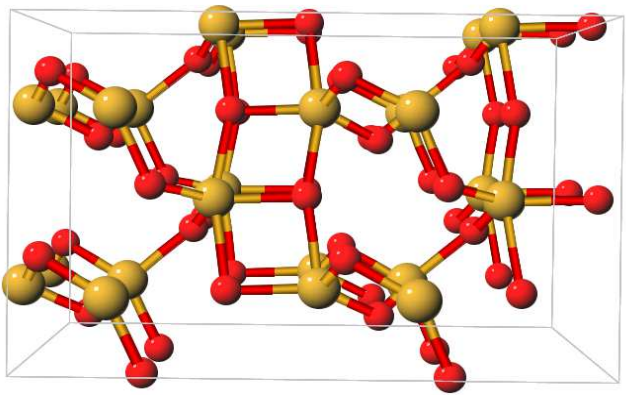}}
\caption{The optimal structure of quartz at the $\Gamma$-point RPBE/STO-3G level of theory.
Note that the picture contains 8 neighboring unit cells for a better 
representation of intercell bonds.}
\label{quartz}
\end{figure}

\begin{table}[h]
\caption{Optimization results for crystal structures
at the PBE/STO-3G level of theory and in the $\Gamma$-point 
approximation, using the QUICCA algorithm.}
\label{optsteps}
\begin{tabular}{lcr}
\toprule
                       & Number of           & Optimum energy\\
Molecule               & optimization steps  &  (a.u.) \\
\colrule
polyethylene           & 8   &    -77.56774   \\
boron-nitride          & 5   &    -78.41368   \\
(10-0)carbon-nanotube  & 7   &  -1503.75024   \\
ice                    & 15  &   -301.31784   \\
quartz                 & 44  &  -1303.03159   \\
sulfur                 & 89  & -12595.53586   \\
\botrule
\end{tabular}
\end{table}

We have developed a periodic test suite in the spirit of Baker's gas phase set \cite{JBaker93}.
The periodic test set includes 6 different systems: Polyethylene, hexagonal boron-nitride, 
a (10,0)carbon-nanotube, hexagonal ice \cite{AGoto90},  quartz \cite{MGTucker01} and sulfur \cite{ACGallacher92}. 
Most of these structures were taken either from the Inorganic Crystal Structures Database
(ICSD) \cite{ICSD} or from Cambridge Crystallographic Data Center
\cite{CCDC} and the translationally unique positions generated with Mercury \cite{Mercury}.  
Details of the geometries used are given in Appendix~\ref{Geometries}.  

Full, simultaneous relaxation of both the lattice and atomic positions have been carried out by means of 
crystal QUICCA, described above, in the $\Gamma$-point approximation at the RPBE/STO-3G level of theory. 

Table \ref{optsteps} shows the number of optimization steps and
the optimal energy for each system. While the first four test systems converged quickly,
quartz and sulfur took substantially longer to reach the optimum.
In the case of quartz, there is a very large (unphysical) deformation, wherein
four membered rings are formed during optimization, due perhaps to a combination of a minimal
basis and $\Gamma$-point effects.  The optimized structure of quartz is shown in Fig.~\ref{quartz}.
In the sulfur crystal, $S_{8}$ rings interact via a Van der Waals like interaction, which has a 
very flat potential, making this a challenging test case.

\subsubsection{Convergence of the energies}

\begin{figure}[h]
\resizebox*{3.5in}{!}{\includegraphics{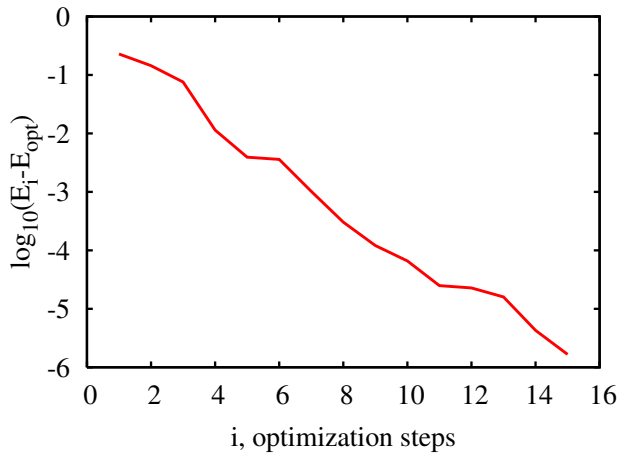}}
\caption{Convergence of the energy during the optimization
of hexagonal ice.}
\label{ICE-energ}
\end{figure}
\begin{figure}[h]
\resizebox*{3.5in}{!}{\includegraphics{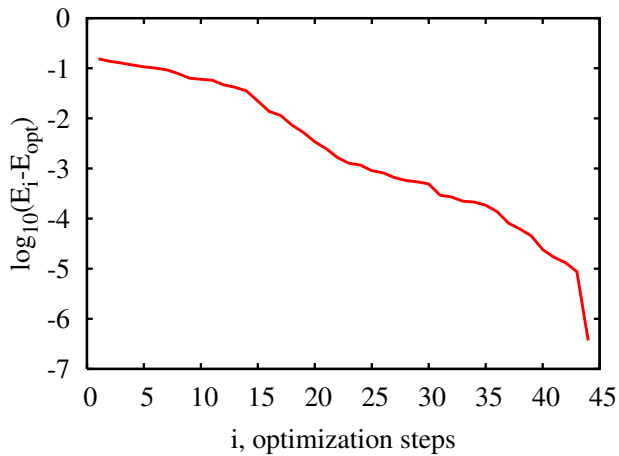}}
\caption{Convergence of the energy during the optimization
of quartz.}
\label{Quartz-energ}
\end{figure}
\begin{figure}[h]
\resizebox*{3.5in}{!}{\includegraphics{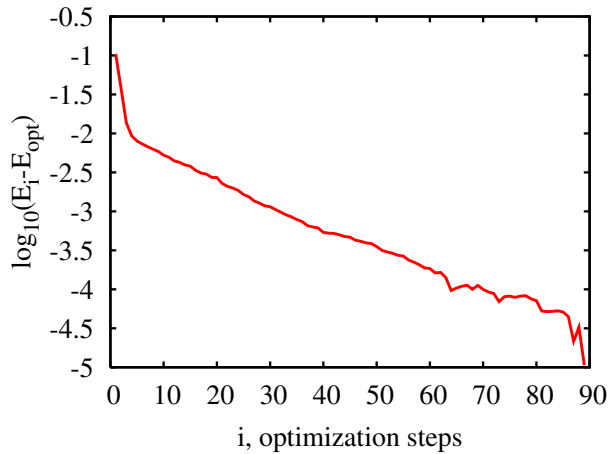}}
\caption{Convergence of the energy during the optimization of sulfur.}
\label{Sulfur-energ}
\end{figure}

Convergence of the total energy is shown in 
Figs.~\ref{ICE-energ}-\ref{Sulfur-energ} for ice, quartz and sulfur.  

\subsubsection{Convergence of the gradients}

\begin{figure}[h]
\resizebox*{3.5in}{!}{\includegraphics{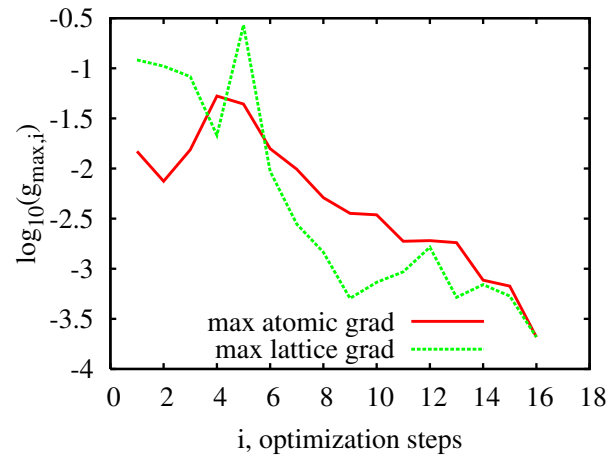}}
\caption{Convergence of the gradients during the optimization 
of hexagonal ice.}
\label{ICE-grads}
\end{figure}

\begin{figure}[h]
\resizebox*{3.5in}{!}{\includegraphics{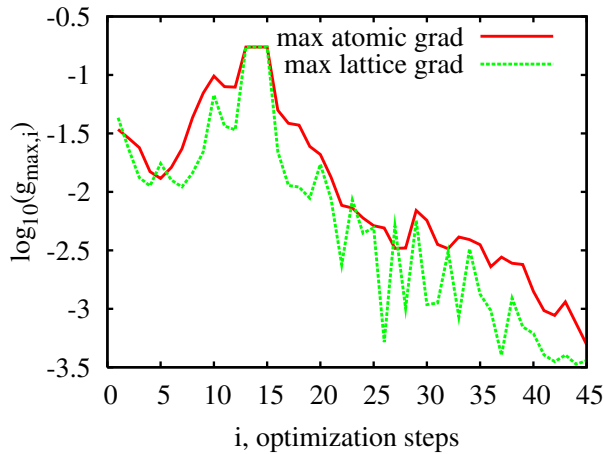}}
\caption{Convergence of the gradients during the optimization of quartz.}
\label{Quartz-grads}
\end{figure}

\begin{figure}[h]
\resizebox*{3.5in}{!}{\includegraphics{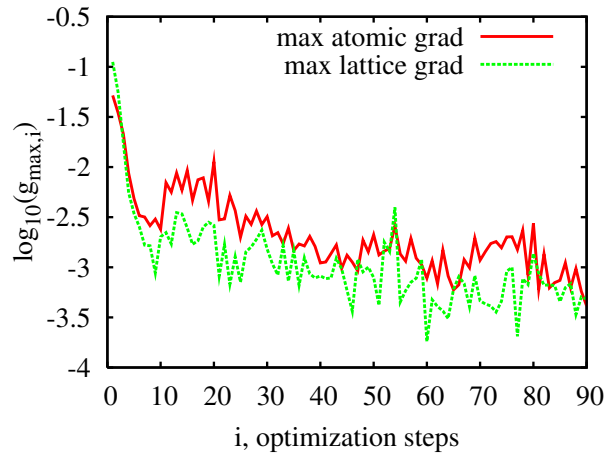}}
\caption{Convergence of the gradients during the optimization of sulfur.}
\label{Sulfur-grads}
\end{figure}

Figures \ref{ICE-grads}-\ref{Sulfur-grads} show convergence
of the maximum Cartesian gradients on atoms and lattice-vector gradients, $g_{{\rm max},i}$,
with optimization step $i$.


\subsection{Urea}

\begin{figure}[h]
\resizebox*{3.5in}{!}{\includegraphics{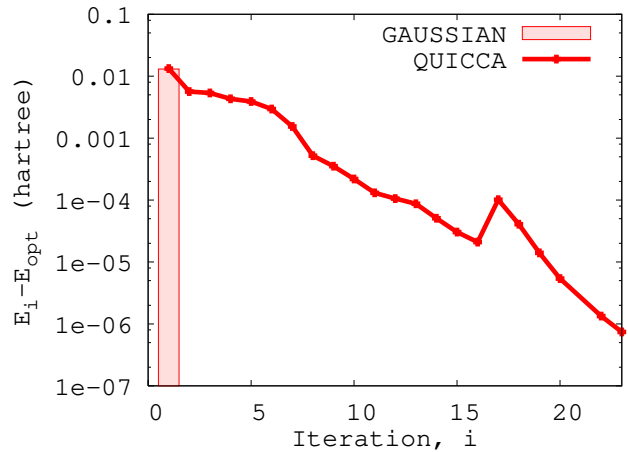}}
\caption{Convergence of the energy during full relaxation of 
RPBE/3-21G urea.  The bar gives the energy difference
computed by Kudin, Scuseria and Schlegel \cite{KKudin01}.}
\label{U321G}
\end{figure}
\begin{figure}[h]
\resizebox*{3.5in}{!}{\includegraphics{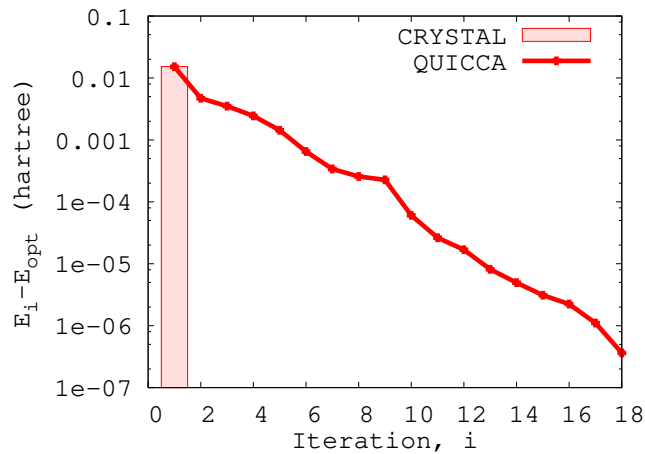}}
\caption{Convergence of the energy during lattice constrained 
optimization of RHF-MIC/STO-3G urea. The bar gives the energy 
difference given by Civalleri {\em et.~al} \cite{BCivalleri01}.}
\label{USTO3G}
\end{figure}

The experimental structure of urea, solved by Swaminanthan {\em et.~al} \cite{SSwaminathan84},
has been used as a benchmark for crystal optimization by several groups.  
Kudin, Scuseria and Schlegel \cite{KKudin01} implemented an early internal coordinate 
optimization scheme in the GAUSSIAN programs, and applied it to the optimization of
RPBE/3-21G urea with internal coordinate constraints.  At about the same time,  
Civalleri {\em et.~al} \cite{BCivalleri01} implemented a Cartesian conjugate gradient 
scheme in the CRYSTAL program, and carried out careful studies examining the effects of 
{\bf k}-space sampling and integral tolerances on fixed lattice optimization of RHF/STO-3G urea.

Here, we make direct contact with these works, Refs.[\onlinecite{KKudin01,BCivalleri01}].
However, because the linear scaling methods used by {\sc MondoSCF} are $\Gamma$-point only, 
we employ a $2\times2\times2$ supercell, so that we may make approximate numerical 
comparison with the {\bf k}-space methods.  This $2\times2\times2$ supercell involves 
16 urea molecules, $C_1$ (no) symmetry, 128 atoms in total, and more than 850 redundant internal 
coordinates;  the number of internal coordinates used by QUICCA fluctuates slightly during optimization.  
For comparison, the work of Civalleri {\em et.~al} \cite{BCivalleri01} makes use of $P\bar{4}2_1m$ 
symmetry, involving just 8 variables in the fixed lattice optimization of urea.  In their 
relaxation of urea, Kudin, Scuseria and Schlegel \cite{KKudin01} 
employed a 4 molecule cell with $S_4$ symmetry, with optimization of the lattice and 
atomic centers, but all dihedral angles constrained, involving 204 redundant internal 
coordinates.

Convergence of the RPBE/3-21G {\sc MondoSCF} calculations are shown in Fig.~\ref{U321G}, 
in which a full relaxation of lattice and atomic centers has been performed, together
with the energy difference from Ref.~[\onlinecite{KKudin01}]. The {\sc GAUSSIAN} values for this 
calculation, involving constrained dihedrals, are {\tt -447.6501595} and {\tt -447.6632120} for the 
beginning and ending values of the total energy.  These (and subsequent) values have been normalized to 
total energy per 2 urea molecules.  The corresponding {\sc MondoSCF} values are {\tt -447.648312} 
and {\tt -447.661578}. The {\sc GAUSSIAN} energy difference is {\tt -0.01305}, while 
the {\sc MondoSCF} difference is {\tt -0.01326}.   The {\sc GAUSSIAN} optimization converged 
in 13 steps, while QUICCA took 24 steps.

Convergence of the RHF-MIC/STO-3G {\sc MondoSCF} calculations are shown in Fig.~\ref{USTO3G}, together
with the energy difference from Ref.~[\onlinecite{BCivalleri01}], both corresponding to 
relaxation of atomic centers only.   The beginning and ending {\sc CYRSTAL} values for this 
calculation are {\tt -442.069368} and {\tt -442.084595}, respectively.  For {\sc MondoSCF},
they are {\tt -442.069473} and {\tt -442.084671}.  The energy differences are {\tt -.01523} and
{\tt -.01520} for {\sc CRYSTAL} and {\sc MondoSCF}, respectively.  The {\sc CRYSTAL} optimization
converged in 15 steps, while QUICCA took 19 steps.

\section{Discussion}\label{discussion}

Overall,  the behavior of QUICCA for crystal systems is similar to that of gas phase systems. 
For a well behaved system like ice, convergence is rapid and monotone. 
For systems undergoing large rearrangements, such as the quartz system, QUICCA
takes many more steps, but still maintains monotone convergence.
For both illconditioned (floppy)  gas phase and crystal systems, such as 
sulfur, convergence is slower, with steps that sometimes raise the energy, 
even with 3-step backtracking.

Large amplitude motions can lead to rapid changes in curvature of the 
potential energy surface.  In this case, the QUICCA algorithm may offer
advantages relative to strategies based on BFGS-like updates, which are 
history laden.   This is because QUICCA employs a weighting scheme that 
takes large moves into account and that can identify recently introduced
trends in just a few steps.

The problems encountered with floppy systems are by no means unique to QUICCA,
but plague most gradient only internal coordinate schemes.  Also, as with other
schemes, we have found the performance of QUICCA to be sensitive to the quality 
of the internal coordinate system.  It is our opinion that the difficulties 
encountered with many floppy systems could be overcome with a better choice
of internal coordinate system.  


For floppy systems, the ability to resolve small energy differences with 
limited precision (due to linear scaling algorithms) can also be problematic.
In particular, with the {\tt TIGHT} option, {\sc MondoSCF} tries to deliver a 
relative precision of 8 digits in the total energy, and 4 digits of absolute 
precision in forces.  For sulfur, achieving atomic forces below {\tt 5D-4} 
corresponds to an energy difference of {\tt 1D-5}, demanding a relative 
precision in the total energy of {\tt 1D-10}.  Exceeding the limits of the
{\tt TIGHT} energy threshold can be seen clearly in Fig.~\ref{Sulfur-energ},
wherein the energy has jumps below {\tt 1D-4}.  The observant reader will also
notice that the $21^{\rm st}$ data point is missing from Fig.~\ref{U321G}.  This 
data point was removed because it was {\tt 1D-4} below the converged value of the 
total energy, {\tt   -3581.2926}, confusing the log-linear plot. These 
fluctuations are at the targeted energy resolution, and are likely due to 
changes in the adaptive HiCu grid for numerical integration of the 
exchange-correlation \cite{CTymczak05a}.  
Nevertheless, reliable structural information can still be obtained, as absolute 
precision in the forces is retained with increasing system size, allowing gradient following
algorithms such as QUICCA to still find a geometry which satisfies the force convergence
criteria. 

For the urea calculations, very good agreement was found between the 
{\sc MondoSCF} calculations and the {\sc CRYSTAL} results, in accordance 
with our previous experience for both pure DFT \cite{CTymczak05a} and RHF-MIC models \cite{CTymczak05b}.  
Slightly less satisfactory agreement was found between the {\sc MondoSCF} and 
{\sc GAUSSIAN} calculations, which was probably due to the differences in 
constraint.  In both cases, the QUICCA calculations took more steps; 
4 more than {\sc CRYSTAL}, and 11 more than {\sc GAUSSIAN}.  It should
be pointed out though, that the {\sc MondoSCF} calculations involved 
a substantially more complicated potential energy surface:  Firstly, 
the $\Gamma$-point surface lacks the symmetry provided by {\bf k}-space
sampling.  Secondly, the $2\times2\times2$ calculation has many more 
degrees of freedom, and in particular, lower frequency modes due to the 
larger cell.  
  
\section{Conclusions} \label{conclusion}

We have implemented the Bu\v{c}ko, Hafner and {\'A}ngy{\'a}n  \cite{TBucko05}
definition of periodic internal coordinates in conjunction with the QUICCA
algorithm, and demonstrated efficient, full relaxation of systems with 
one, two and three dimensional periodicity.  In general, we have found that
QUICCA performs with an efficiency comparable to that of similarly posed gas 
phase problems, and speculate that further enhancement  may be achieved through
an improved choice of internal coordinates.

We have argued that linear scaling internal coordinate transformations for 
crystal systems can be achieved with a left-handed pseudo-inverse, as the 
dense rows and columns of the periodic ${\bf B}^t {\bf B}$ matrix determine
just the last few pivots of the corresponding Cholesky factor.

We have carried out supercell calculations using a full compliment of
linear scaling algorithms, including sparse linear algebra, fast force and 
matrix builds and found good agreement with {\bf k}-space methods, involving 
a modest number of optimization cycles.  Thus, in addition to further demonstrating 
the stability of our linear scaling algorithms, we have established QUICCA as a 
reliable tool for large scale optimization problems in the condensed phase.

In conclusion, QUICCA is a new gradient only approach to internal coordinate
optimization that is robust and generally applicable, both to gas-phase
molecules and systems of one, two and three dimensional periodicity.  It 
allows for flexible optimization protocols, involving simultaneous 
relaxation of lattice and atom centers, constrained lattice with relaxation of 
atom centers, constrained atom centers with optimization of the lattice,
admixtures of the above with constrained internal coordinates, etc. 
QUICCA is conceptually simple and easy to implement.
Perhaps most importantly though, it is a new approach to gradient only 
internal coordinate optimization, offering a number of opportunities for 
further development.

\begin{acknowledgments}
 This work has been supported by the US Department of Energy 
 under contract W-7405-ENG-36 and the ASCI project.  The authors
 thank C.~J.~Tymczak, Valery Weber and Anders Niklasson for helpful 
 comments.
\end{acknowledgments}

\bibliography{mondo_new}

\appendix

\section{TEST SET COORDINATES}\label{Geometries}

Here, input geometries for the crystal optimization test suite are detailed.  These
geometries are available as supplementary data, and are also available from the
authors upon request.

\paragraph{Polyethylene}
The 1-D periodic structure of polyethylene is given in Table~\ref{PE}.
\begin{table}[h]
\caption{Atomic coordinates in $\AA$ for the elementary unit cell of polyethylene,
with corresponding lattice parameters $a=2.0\AA, b=c=0.00, \alpha=\beta=\gamma=90.0^\circ$.}\label{PE}
\begin{tabular}{lrrr}
\hline
\tt C  & \tt  0.500 & \tt  0.500 & \tt  0.000 \\
\tt H  & \tt  0.500 & \tt  1.300 & \tt  0.800 \\
\tt H  & \tt  0.500 & \tt  1.300 & \tt -0.800 \\
\tt C  & \tt  1.500 & \tt -0.500 & \tt  0.000 \\
\tt H  & \tt  1.500 & \tt -1.300 & \tt  0.800 \\
\tt H  & \tt  1.500 & \tt -1.300 & \tt -0.800 \\
\hline
\end{tabular}
\end{table}
\paragraph{Hexagonal boron-nitride}
The 2-D periodic coordinates for hexagonal boron-nitride are given in Table~\ref{BN}.
\begin{table}[h]
\caption{Fractional coordinates for the elementary unit cell of hexagonal boron-nitride 
with corresponding lattice parameters $a=b=2.420\AA, c=0.00, \alpha=\beta=90.0^\circ,$ 
and $\gamma=120.0^\circ$.}\label{BN}
\begin{tabular}{lrrr}
\hline
\tt B &0.33333333333 &0.1666666666 &0.00 \\
\tt N &0.66666666666 &0.8400000000 &0.00 \\
\hline
\end{tabular}
\end{table}
\paragraph{(10,0)carbon-nanotube}
The geometry of the 1-D periodic (10,0)carbon-nanotube has 
all bond-lengths parallel to the nanotube axis initially at $1.480${\AA}, 
while those running perpendicular to the axis are $1.402${\AA} long.
The lattice length is $a=4.44${\AA}, with the elementary cell containing 
40 atoms. While this data entirely determines the structure of the symmetric 
(10,0)carbon-nanotube, it is also available as supplementary data.

\paragraph{Ice}
Hexagonal ice is the most important natural occurrence of ice.
Its structure has been taken from the literature \cite{AGoto90}. 
Since the literature provides two equilibrium position
for each hydrogen atom, due to the tunneling of hydrogens in ice,
our starting structure is taken as the average of these two positions
for each hydrogen atom, and is given in Table~\ref{ICE}.
\begin{table}[h]
\caption{Atomic coordinates in $\AA$ for the elementary unit cell of hexagonal ice,
with corresponding lattice parameters $a=b=4.511 \AA, c=7.346 \AA, \alpha=90.0^\circ, 
\beta=90.0^\circ$ and $\gamma=120.0^\circ$
\label{ICE}
}
\begin{tabular}{lrrr}
\hline
\tt  O & \tt   0.0000   &   \tt  2.6040000  & \tt    3.216 \\
\tt  H & \tt   2.2555   &   \tt  0.0003594  & \tt    3.673 \\
\tt  H & \tt   1.1280   &   \tt  1.9540000  & \tt    3.673 \\
\tt  O & \tt   2.2560   &   \tt  1.3020000  & \tt    4.130 \\
\tt  H & \tt  -1.1280   &   \tt  1.9540000  & \tt    3.673 \\
\tt  H & \tt   2.2560   &   \tt  1.3020000  & \tt    5.510 \\
\tt  O & \tt   2.2560   &   \tt  1.3020000  & \tt    6.889 \\
\tt  H & \tt   1.1280   &   \tt  1.9540000  & \tt    7.346 \\
\tt  H & \tt  -1.1280   &   \tt  1.9540000  & \tt    7.346 \\
\tt  O & \tt   0.0000   &   \tt  2.6040000  & \tt    7.803 \\
\tt  H & \tt   0.0000   &   \tt  2.6040000  & \tt    9.183 \\
\tt  H & \tt   2.2555   &   \tt  0.0003594  & \tt    7.346 \\
\hline
\end{tabular}
\end{table}
\paragraph{Quartz}
The initial structure of quartz was taken from Ref.~[\onlinecite{MGTucker01}],
and has 9 atoms in the unit cell.
\begin{table}[h]
\caption{Atomic coordinates in $\AA$ for the elementary unit cell of quartz,
with corresponding lattice parameters 
$a=b=4.9019 {\AA}, c=5.3988 {\AA}, \alpha=\beta=90.0^\circ $ 
and $\gamma=120.0^\circ $.}\label{QTZ}
\begin{tabular}{rrrr}
\hline
\tt  Si  &   \tt   1.306   &  \tt   2.261  &   \tt   0.000  \\
\tt   O  &   \tt   2.768   &  \tt   2.492  &   \tt   4.772  \\
\tt  Si  &   \tt  -1.145   &  \tt   1.984  &   \tt   3.599  \\
\tt   O  &   \tt   1.360   &  \tt   1.151  &   \tt   1.172  \\
\tt   O  &   \tt   3.225   &  \tt   0.602  &   \tt   2.972  \\
\tt   O  &   \tt   0.317   &  \tt   1.753  &   \tt   4.226  \\
\tt  Si  &   \tt   2.291   &  \tt   0.000  &   \tt   1.800  \\
\tt   O  &   \tt  -1.091   &  \tt   3.094  &   \tt   2.427  \\
\tt   O  &   \tt   0.774   &  \tt   3.643  &   \tt   0.627  \\
\hline
\end{tabular}
\end{table}
\paragraph{Sulfur}
The structure of sulfur was taken from Ref.~[\onlinecite{ACGallacher92}].
Containing 32 atoms, this structure is available as supplementary data.

\end{document}